\documentclass[english,prl,twocolumn,showpacs,amsmath,amssymb,superscriptaddress,floatfix]{revtex4}
\usepackage[T1]{fontenc}
\usepackage[latin1]{inputenc}
\usepackage{graphicx}
\makeatletter

\usepackage{bm}
\usepackage{pifont}

\usepackage{babel}
\makeatother

\begin{document}


\title{Quantized Charge Pumping through a Carbon Nanotube Double Quantum Dot}

\author{S.J. Chorley}
\author{J. Frake}
\author{C.G. Smith}
\author{G.A.C. Jones}
\author{M.R. Buitelaar}\email{mrb51@cam.ac.uk}
\affiliation{Cavendish Laboratory, University of Cambridge, Cambridge CB3 0HE, United Kingdom}

\begin{abstract}
We demonstrate single-electron pumping in a gate-defined carbon
nanotube double quantum dot. By periodic modulation of the potentials of
the two quantum dots we move the system around charge
triple points and transport exactly one electron or hole per cycle. We investigate the pumping as a function of the modulation
frequency and amplitude and observe good current quantization up to frequencies of 18 MHz
where rectification effects cause the mechanism to break down.
\end{abstract}





\maketitle

Fast and accurate control of the charge state of quantum dots is
important in research areas varying from quantum information
processing to quantum metrology. Quantum computation schemes based
on charge or spin qubits defined in quantum dots \cite{Loss}, for
example, require manipulation of the charge state well within the
coherence times of the qubits which, in practise, implies nanosecond
control. In quantum metrology, efforts towards the development of a
current standard require the transfer of single electrons at
nanosecond timescales with a precision of 0.1 part per million or
better \cite{Zimmerman}. The ability of fast control of the charge
states of carbon nanotube quantum dots is of particular interest in
these respects. In quantum information processing, carbon nanotube
quantum dots are attractive as electron spin coherence times are
expected to be long and because spin-orbit interaction
\cite{Kuemmeth, Jespersen, Chorley1} allows for electrical or even
optical \cite{Galland} control of the spin states. Carbon nanotube
quantum dots are also promising as single electron pumps in quantum
metrology for use in a current standard. The ultimate precision of
single-electron pumps is widely regarded to be dependent on the
strength of electron-electron interactions, suppressing errors due
to co-tunneling events \cite{Jensen}. These interactions are
exceptionally strong in carbon nanotubes. In addition, when combined
with electron-hole recombination, the ability to transfer single
electrons in semiconducting nanotubes at well defined intervals has
potential in quantum optics as an electrically-driven on-demand
single-photon source in the infrared frequency range \cite{Mueller,
Hogele}.

Here we demonstrate single-electron pumping in a gate-defined carbon
nanotube double quantum dot by periodic modulation of the potentials
of the two quantum dots around charge triple points in the stability
diagram. We investigate the pumping as a function of the modulation
frequency and amplitude and show quantized charge pumping up to
frequencies of 18 MHz, corresponding to a period of $\sim 55$
nanoseconds. The device we consider is a single-walled carbon
nanotube grown by chemical vapour deposition on degenerately doped
Si terminated by 300 nm SiO$_2$, see Fig.~1(a). To define a double
quantum dot, the nanotube is contacted by evaporated source and
drain electrodes of Ti/Au which form the outer tunnel barriers of
the quantum dots. A capacitively coupled top gate, separated from
the nanotube by $\sim 3$ nm of AlO$_x$ is used to control the tunnel
coupling between the dots. The outer two plunger gates, set back
from the nanotube, control the electron number on each dot.

\begin{figure}
\includegraphics[width=84mm]{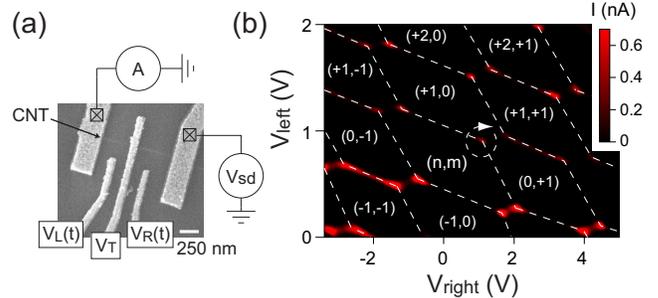}
\caption{\label{Fig1} (color online) \textbf{(a)} Scanning electron
micrograph of a typical carbon nanotube (CNT) device with simplified
electrical setup. \textbf{(b)} Charge stability diagram of the carbon
nanotube double quantum dot for $V_{sd} = 0.5$ mV measured at $T=40$ mK.
The ordered pairs $(n,m)$ represent the excess electrons on the two dots. The circle shows a typical
pumping trajectory around a triple point.}
\end{figure}

The device was bonded onto microwave printed circuit board and
mounted on the tail of a $40\,$mK dilution refrigerator. The two
plunger gates are connected to semirigid coaxial cables via bias tees
at the mixing chamber, allowing dc and rf signals to be applied. The
source is connected to a resonant circuit with dc connection to
allow simultaneous transport and rf reflectometry measurements
\cite{Chorley2}. High frequency synthesized waveforms were applied
to the gates using a Tektronix AWG5014 arbitrary waveform generator
and the current through the nanotube device is measured by a
Keithley 6514 electrometer.

\begin{figure*}
\includegraphics[width=165mm]{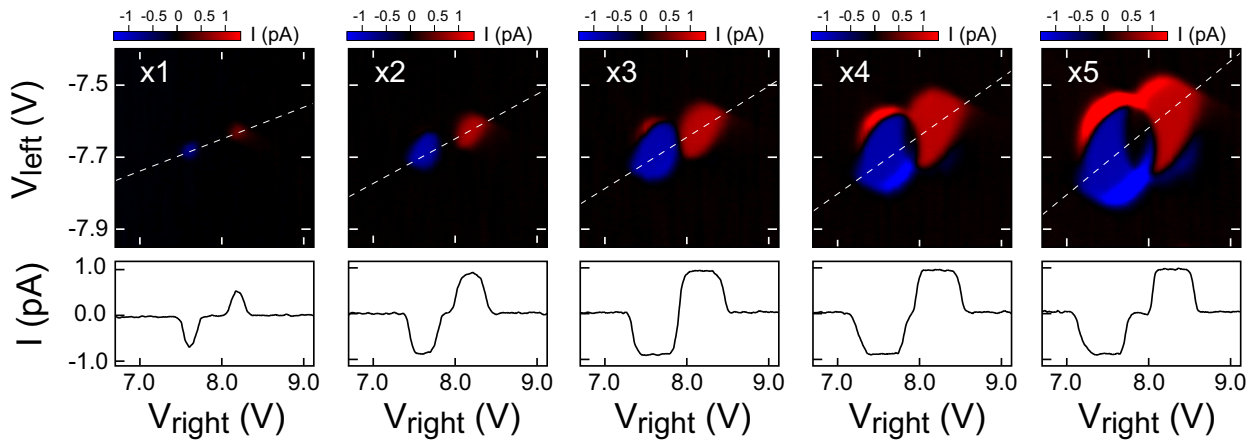}
\caption{\label{Fig2} (color online) Pumped current stability
diagram as a function of gate oscillation amplitude. For the
leftmost measurement, the amplitude of the oscillation voltages on
$V_L$ and $V_R$ are 25 and 87.5 mV, respectively \cite{amplitude}.
The applied potentials are then increased in steps as indicated,
keeping the ratio fixed. The applied bias $V_{sd} = 0$. The line
traces show the current for the paths through the electron and hole
pumping cycles, as indicated in the top panels by the dashed lines.
The current is quantized for both polarities.}
\end{figure*}

At low temperatures, quantum dots form between each contact and the
central barrier. For appropriate settings of the Si back gate
voltage $V_{BG}=1$ V and top gate voltage $V_T = -0.05$ V, the
charge stability diagram displays the characteristic honeycomb
pattern of a double quantum dot \cite{Wiel,Graber}, see Fig.~1(b). The
electron occupation number of the quantum dots is indicated by the
ordered pairs $(n,m)$. A finite conductance is observed at the
triple points where three different charge states are degenerate.
Weak co-tunneling is observed for some charge transitions of the
left quantum dot but mostly suppressed otherwise. The charging
energies of the two quantum dots can be obtained from the charge
stability diagrams and yield $U_L \approx 6$ and $U_R \approx 5$ meV
for the left and right quantum dot, respectively. The interdot
charging energy $U' \approx 1.3$ meV.

\begin{figure}[b]
\includegraphics[width=82mm]{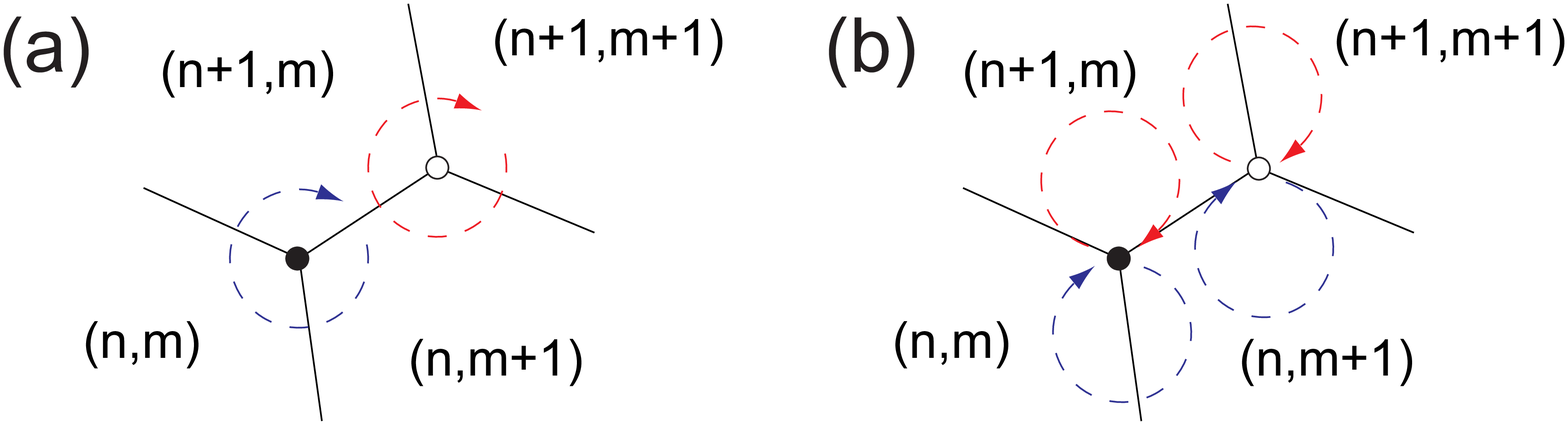}
\caption{\label{Fig3}(color online) \textbf{(a)} Pumping trajectories around the triple
points in the double dot stability diagram. The color indicates the polarity of the current which is opposite
for the two triple points. \textbf{(b)} Pumping trajectories that cross the triple points. The color indicates the polarity of the current due to rectification.}
\end{figure}

Following a technique first demonstrated for metallic
single-electron tunneling devices \cite{Pothier}, the occupation
number $(n,m)$ of the quantum dots is varied periodically by
applying a small sinusoidal voltage to each plunger gate with a 90$^\circ$
phase shift between them such that the gates trace out
approximately circular paths on the stability diagram \cite{amplitude}. The centers
of the circular paths are then varied by addition of a dc signal
offset while the source-drain bias $V_{sd}$ is kept at zero. Figure 2 shows the resulting current around a representative triple point pair of our nanotube double quantum
dot for a modulation frequency $f=6$ MHz and a sequence of increasing amplitude. The line traces show the current along the
dashed lines in the top panels for each measurement. For the lowest pumping amplitude, current peaks of opposite
polarity are observed at the two triple points and the current is
zero elsewhere. As the amplitude of the signal is increased, clear
current plateaus of about 0.96 pA are observed. The plateaus are
flat within our measurement accuracy of $\Delta I \sim 50$ fA.

These observations can be understood considering the pumping sequence illustrated in Fig.~3(a).
When a path on the stability diagram encircles a triple point, exactly one electron or hole
moves between the electrodes per cycle and the current is expected
to be quantized as $I=ef$, where $e$ is the electron charge \cite{Pothier,Fuhrer}. For the leftmost trajectory in
Fig.~3(a), for example, electrons are moved in the sequence
$(n,m)\rightarrow (n+1,m)\rightarrow (n,m+1)\rightarrow (n,m)$. When
the same path is encircling the other triple point of a pair, the
polarity of the current is reversed. For the modulation frequency $f=6$ MHz used in Fig.~2,
this corresponds to the observed currents of 0.96 pA. The area in the stability diagram in which a quantized current is
observed increases with pumping amplitude up to a maximum set by
$U'$. As expected, the magnitude of the current observed on the plateaus in Fig.~2 does not increase
with pumping amplitude and depends on the modulation frequency only. With increasing pumping amplitudes (rightmost plots), there are
trajectories that include both triple points such that no charge is
transferred between the dots but only between the dots and leads,
i.e., following a sequence $(n,m)\rightarrow (n+1,m) \rightarrow
(n+1,m+1) \rightarrow (n,m+1) \rightarrow (n,m)$. This results in an
area of zero current in between the plateaus.

In the rightmost plots of Fig.~2, another effect becomes visible where non-quantized currents are observed above and below
the quantized regions. These positions in the stability diagram
correspond to those trajectories that cross the triple points where
the double dot has a finite conductance and a non-quantized current
flows when $V_{sd} \neq 0$. The effect can therefore be understood
as rectification where modulation of the source and drain potentials
by the plunger gates - to which they are capacitively coupled -
induces a current. The polarity of the rectified currents depends on
the direction of the trajectories in the stability diagram near the
triple points \cite{clockwise}, as indicated in Fig.~3(b).
Rectification effects become stronger for larger amplitudes and
frequencies preventing us from observing a clean quantized current
above $\sim 18$ MHz.

These observations are further illustrated by Fig.~4 which shows the
current at the triple points for increasing frequencies. The current
follows the expected relation $I=ef$, indicated by the dashed lines,
upto 18 MHz where it starts to deviate \cite{phase} as rectification
effects become dominant. We note that 18 MHz is a much smaller
frequency than the intrinsic dynamic limit of the device which is
given by its $RC$ time constant \cite{Jensen,Reilly} and which we
estimate to be of order 1-10 GHz \cite{estimate}. This suggests that
current quantization could be observed for much larger modulation
frequencies if rectification can be sufficiently suppressed. Indeed,
in previous work in which potential modulation of nanotube quantum
dots was achieved by surface-acoustic-waves (SAW) - avoiding direct
coupling between the rf source and nanotube contacts - quantized
currents in the GHz range were observed \cite{Buitelaar,Wurstle}. We
believe rectification can be reduced in the present double dot
devices by a different choice of substrate, e.g., undoped Si or
quartz, and stronger coupling between the quantum dots and plunger
gates.

\begin{figure}
\includegraphics[width=82mm]{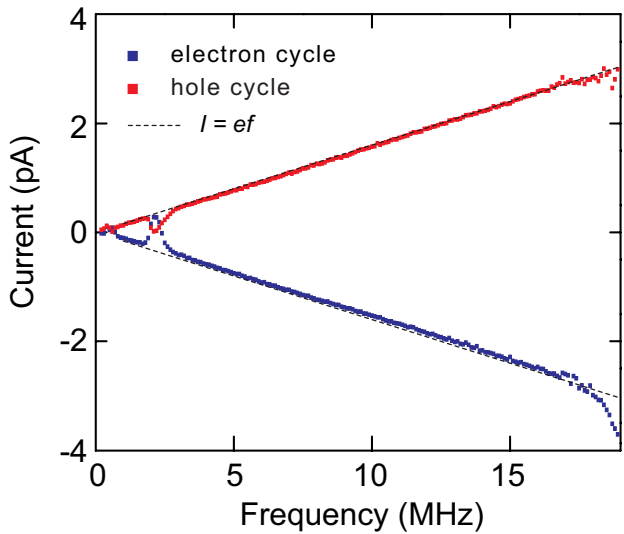}
\caption{\label{Fig4} (color online) Pumped current as a function
of frequency for the electron (blue) and hole cycle (red) for clockwise trajectories
in the stability diagram. The
amplitude of the gate modulation is as used in the middle panel of
Fig.~2. The dashed lines correspond to $I=\pm ef$.}
\end{figure}

In conclusion, we demonstrate charge pumping in a carbon nanotube double quantum
by periodic modulation of the dot potentials. We investigate the pumping as a function
of the modulation amplitude and frequency and observe quantized currents up to
frequencies of $\sim 18$ MHz. Above this frequency, rectification prevents us from
observing the quantization cleanly. We believe that for an optimized
device geometry accurately quantized currents in the GHz range are experimentally feasible.

We thank David Cobden and Jiang Wei for the carbon nanotube growth and Karl Petersson,
Victoria Russell, and Mamta Thangaraj for technical assistance.
This work was supported by EPSRC, the Newton Trust and the Royal
Society (M.R.B.).


\begin{thebibliography}{10}

\bibitem{Loss}
D. Loss and D.P. DiVincenzo, Phys. Rev. A \textbf{57}, 120 (1998).

\bibitem{Zimmerman}
N.M. Zimmerman, Physics Today \textbf{63}, 68 (2010).

\bibitem{Kuemmeth}
F. Kuemmeth, S. Ilani, D.C. Ralph, and P.L. McEuen, Nature (London)
\textbf{452}, 448 (2008).

\bibitem{Jespersen}
T.S. Jespersen, K. Grove-Rasmussen, J. Paaske, K. Muraki, T.
Fujisawa, J. Nyg{\aa}rd, and K. Flensberg, Nature Phys. \textbf{7},
348 (2011).

\bibitem{Chorley1}
S.J. Chorley, G. Giavaras, J. Wabnig, G.A.C. Jones, C.G. Smith,
G.A.D. Briggs, and M.R. Buitelaar, Phys. Rev. Lett. \textbf{106},
206801 (2011).

\bibitem{Galland}
C. Galland and A. Imamo\v{g}lu, Phys. Rev. Lett. \textbf{101},
157404 (2008).

\bibitem{Jensen}
H.D. Jensen and J.M. Martinis, Phys. Rev. B \textbf{46}, 13407
(1992).

\bibitem{Mueller}
T. Mueller, M. Kinoshita, M. Steiner, V. Perebeinos, A.A. Bol, D.B.
Farmer, and P. Avouris, Nature Nanotech. \textbf{5}, 27 (2010).

\bibitem{Hogele}
A. Hogele, C. Galland, M. Winger, A. Imamoglu, Phys. Rev. Lett.
\textbf{100}, 217401 (2008).

\bibitem{Chorley2}
S.J. Chorley, J. Wabnig, Z.V. Penfold-Fitch, K.D. Petterson, J.
Frake, C.G. Smith, and M.R. Buitelaar, Phys. Rev. Lett.
\textbf{108}, 036802 (2012).

\bibitem{Wiel}
W.G. van der Wiel, S. De Franceschi, J.M. Elzerman, T. Fujisawa, S.
Tarucha, and L.P. Kouwenhoven, Rev. Mod. Phys. \textbf{75}, 1
(2002).

\bibitem{Graber}
M.R. Gr\"{a}ber, W.A. Coish, C. Hoffmann, M. Weiss, J. Furer, S.
Oberholzer, D. Loss, and C. Sch\"{o}nenberger, Phys. Rev. B
\textbf{74}, 075427 (2006).

\bibitem{Pothier}
H. Pothier, P. Lafarge, C. Urbina, D. Esteve, and M.H. Devoret,
Europhys. Lett. \textbf{17}, 249 (1992).

\bibitem{amplitude}
The amplitude of the two output signals at the waveform generator is tuned such
that the potential modulation at the device is of similar strength for both dots.
Since the right gate is more weakly coupled to the right dot as
compared to the coupling between the left plunger gate and left
dot by a factor $C_L/C_R \sim 3.5$, the voltage we applied to the
right plunger is correspondingly larger.

\bibitem{Fuhrer}
A. Fuhrer, C. Fasth, and L. Samuelson, Appl. Phys. Lett.
\textbf{91}, 052109 (2007).

\bibitem{clockwise}
We verified that both the quantized and non-quantized currents
reverse polarity as the direction of the trajectories is changed
from clockwise to anti-clockwise.

\bibitem{phase}
At $f \sim 2.5$ MHz, a resonance in one of the plunger gate's transmission lines changes the phase angle
and power of the signal, resulting in a change of the polarity of the pumped current.

\bibitem{Reilly}
D.J. Reilly and T.M. Buehler, Appl. Phys. Lett. \textbf{87}, 163122
(2005).

\bibitem{estimate}
Here we use a resistance $R \sim 2$ M$\Omega$ and capacitance $C \sim 20$ aF as deduced from
the dc transport measurements.

\bibitem{Buitelaar}
M.R. Buitelaar, P.J. Leek, V.I. Talyanskii, C.G. Smith, D. Anderson,
G.A.C. Jones, J. Wei, and D.H. Cobden, Semicond. Sci. Technol.
\textbf{21}, S69 (2006).

\bibitem{Wurstle}
C. W\"{u}rstle, J. Ebbecke, M.E. Regle, and A. Wixforth, New Journal
of Physics \textbf{9}, 73 (2007).

\end{thebibliography}
\end{document}